\documentclass[usenatbib]{mn2e}
\usepackage{txfonts}
\usepackage{graphicx}
\usepackage{array}
\usepackage{longtable}
\bibpunct{(}{)}{;}{a}{}{,}
\def\z{\lower1.2pt\hbox{*}}
\def\zz{\lower1.2pt\hbox{**}}

\title[Young open cluster NGC 7129] {The distance to the young cluster
NGC 7129 and its age}

\author[V. Strai\v{z}ys et al.]
{V. Strai\v{z}ys,$^{1}$\thanks{E-mail:vytautas.straizys@tfai.vu.lt}
M. Maskoli\={u}nas,$^{1}$
R. P. Boyle,$^{2}$ P. G. Prada Moroni,$^{3,4}$ E. Tognelli,$^{3}$
\newauthor K. Zdanavi\v{c}ius,$^{1}$ J. Zdanavi\v{c}ius,$^{1}$
V. Laugalys$^{1}$ and A. Kazlauskas$^{1}$\\
\\
     $^{1}$~Institute of Theoretical Physics and Astronomy, Vilnius
University, Go\v{s}tauto 12, Vilnius LT-01108, Lithuania\\
    $^{2}$~Vatican Observatory Research Group, Steward Observatory,
 Tucson, AZ 85721, U.S.A.\\
    $^{3}$~Physics Department ``E. Fermi'', University of Pisa, Largo
B. Pontecorvo 3, I-56127 Pisa, Italy\\
    $^{4}$~INFN, Sezione di Pisa, Largo B. Pontecorvo 3, I-56127 Pisa,
Italy}

\begin{document}

\date{Accepted 2013 November 30.  Received 2013 November 28; in original
form 2013 October 17} \pagerange{1--8}
\pubyear{2013}

\maketitle

\label{firstpage}

\begin{abstract} The dust cloud TGU H645 P2 and embedded in it young
open cluster NGC 7129 are investigated using the results of medium-band
photometry of 159 stars in the Vilnius seven-colour system down to $V$ =
18.8 mag.  The photometric data were used to classify about 50\,\% of
the measured stars in spectral and luminosity classes.  The extinction
$A_V$ vs. distance diagram for the 20$\arcmin$\,$\times$\,20$\arcmin$
area is plotted for 155 stars with two-dimensional classification from
the present and the previous catalogues.  The extinction values found
range between 0.6 and 3.4 mag.  However, some red giants, located in the
direction of the dense parts of the cloud, exhibit the infrared
extinction equivalent up to $A_V$ = 13 mag.  The distance to the cloud
(and the cluster) is found to be 1.15 kpc (the true distance modulus
10.30 mag).  For determining the age of NGC 7129, a luminosity vs.
temperature diagram for six cluster members of spectral classes B3 to A1
was compared with the Pisa pre-main-sequence evolution tracks and the
Palla birthlines.  The cluster can be as old as about 3 Myr, but star
forming continues till now as witnessed by the presence in the cloud of
many younger pre-main-sequence objects identified with photometry from
2MASS, Spitzer and WISE infrared surveys.  \end{abstract}


\section{Introduction}

A small group of early-type stars surrounded by the
reflection nebula NGC 7129 was described by \citet{Herbig1960} in
his pioneering paper on the search of emission-line B and A stars
associated with nebulosities. Later on, more fainter stars in the
nebula and around it were identified \citep{Strom1976, Hartigan1985,
Hodapp1994}.
The cluster was included in the catalogue of infrared
star clusters and stellar groups by \citet*{Bica2003a}.  The dust cloud,
now known as TGU H645 P2 \citep{Dobashi2005}, is illuminated mainly by
the three hot stars:  BD+65 1637 (B2e, $V$ = 10.15), BD+65 1638 (B3, $V$
= 10.18) and LkH$\alpha$\,234 (B6e, $V$ = 11.90); the 1st and the 3rd of
them are the Herbig Ae/Be stars \citep{Herbig1960}.  Most fainter stars
of the cluster are also young stellar objects (hereafter YSOs)
identified by \citet{Magakian2004} using spectral observations and {\it
VRI} photometry, and by \citet{Gutermuth2004, Gutermuth2009},
\citet{Muzerolle2004}, \citet{Stelzer2009} using 2MASS and Spitzer
infrared photometry and Chandra X-ray survey.  \citet{Kato2011}
identified some possible YSOs in the close vicinity of LkH$\alpha$\,234
with high-resolution near IR images.  For a few YSOs in the cluster and
in the dark cloud LDN 1181 located about 0.5$\degr$ north of NGC 7129,
masses, spectral types, emission-line intensities, {\it BVRI}
photometry, spectral energy distributions and physical parameters were
obtained by \citet{Kun2009}.  A review of the investigations of NGC 7129
and its individual objects was published by \citet{Kun2008}.

The distance determinations to NGC 7129 are scarce and of low accuracy.
The first distance estimation was done by \citet{Racine1968}, who used
one-dimensional spectral classes of the stars BD+65 1637 and BD+65 1638
(B2nne and B3) and their {\it UBV} photometry.  The distance moduli 12.2
(2.75 kpc) and 10.0 (1 kpc) were estimated, but they are very uncertain,
since the luminosity classes of these stars were unknown.  The last of
these two distance values has been used in most subsequent
investigations, including the recent papers in the Spitzer era.  The
next estimate of the distance to the NGC 7129 complex, 1.26 kpc, was
published by \citet{Shevchenko1989} using {\it BV} photometry and
spectral classification of a few reddened stars.  However, their result
is also of low accuracy since the authors have used one-dimensional
spectral classes estimated visually from low-dispersion objective prism
spectra.  Thus, NGC 7129 remains one of the rare NGC objects to which
the distance and age lines in the WEBDA database are empty.

The stars in the direction of NGC 7129 exhibit very different reddenings
and extinctions making the traditional method for determining the
cluster parameters by fitting ZAMS to the reddened main sequence not
applicable.  Thus we need to make individual search of its members,
their classification and determining colour excesses and extinctions.
And what is more, because of young age only a few cluster members of
spectral class B in the colour-magnitude diagram are located close to
the main sequence.  The majority of stars are on the pre-main-sequence
(hereafter PMS) evolutionary tracks, thus their spectral classification
and reddening determination are of low quality.

With the aim to determine the distance to the dust cloud TGU H645 P2 and
the embedded cluster NGC 7129, we decided to apply the Vilnius
seven-colour system.\footnote{~The mean wavelengths of the passbands are
345 ($U$), 374 ($P$), 405 ($X$), 466 ($Y$), 516 ($Z$), 544 ($V$) and 656
($S$) nm.  For more details about the system see in the monograph
\citet{Straizys1992}, available in the pdf format from//
http://www.itpa.lt/MulticolorStellarPhotometry/} This system makes
possible to classify normal stars of various temperatures in spectral
and luminosity classes in the presence of variable interstellar
reddening.  In this area, the first paper on Vilnius photometry was
published by \citet{Maskoliunas2012}, hereafter Paper I. It contains the
results of CCD photometry of 2140 stars down to $V$ = 17 mag in the 1.5
square degree area, which covers the clusters NGC 7129 and NGC 7142, and
a surrounding area.  The results of photometric classification in
spectral and luminosity classes were given for about 70\% of the
measured stars.  The cluster NGC 7142 has been investigated in our
previous paper \citep{Straizys2013}, hereafter Paper II.

To increase the limiting magnitude, we have obtained CCD exposures
with a larger telescope and a better resolving power. This paper
describes the new results of photometry and classification of stars
in the direction of NGC 7129 and their application to determine more
reliable distance and age of the cluster.

\section{Observations, their processing and results}


\begin{figure}
\resizebox{\hsize}{!}{\includegraphics{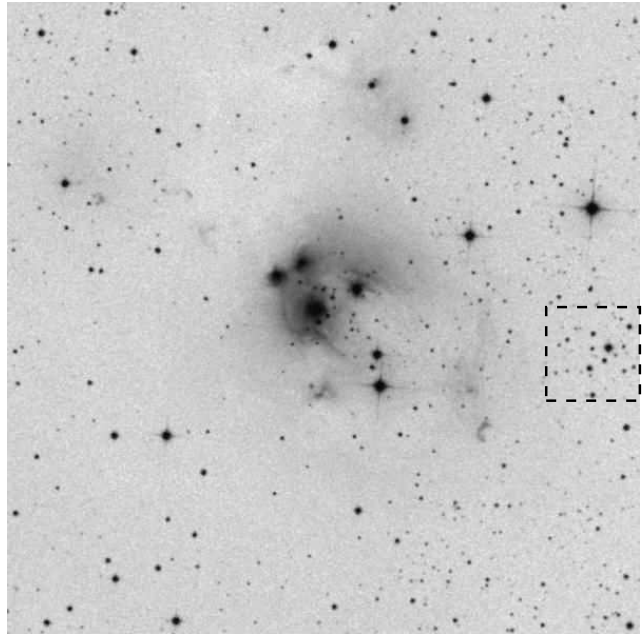}}
\vskip1mm
\caption{Area of the cluster NGC 7129 observed in the Vilnius
photometric system with VATT (13$\arcmin$\,$\times$\,13$\arcmin$).  The
2$\arcmin$\,$\times$\,2$\arcmin$ square surrounds a group of stars which
was suspected by \citet{Bica2003b} as an infrared cluster.  More about
this group see in Section 7. The DSS2 Red map from SkyView.}

\label{1}
\end{figure}


\begin{figure}
\resizebox{\hsize}{!}{\includegraphics{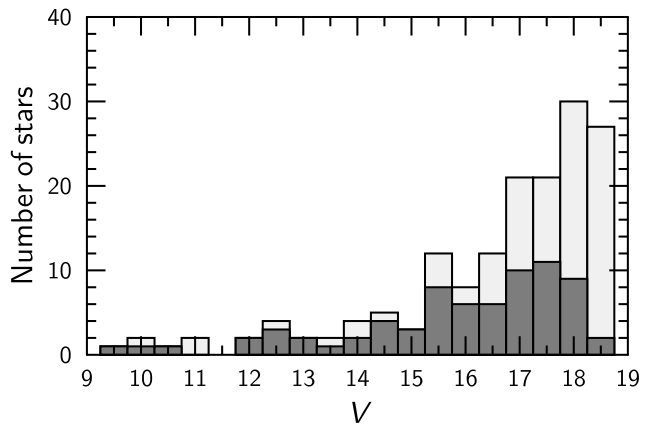}}
\vskip1mm
 \caption{Distribution of the measured stars in the NGC 7129 area
in apparent magnitudes. The shadowed parts of the columns correspond to
stars for which two-dimensional spectral types were determined.}
\label{2}
\end{figure}


\begin{figure}
\resizebox{\hsize}{!}{\includegraphics{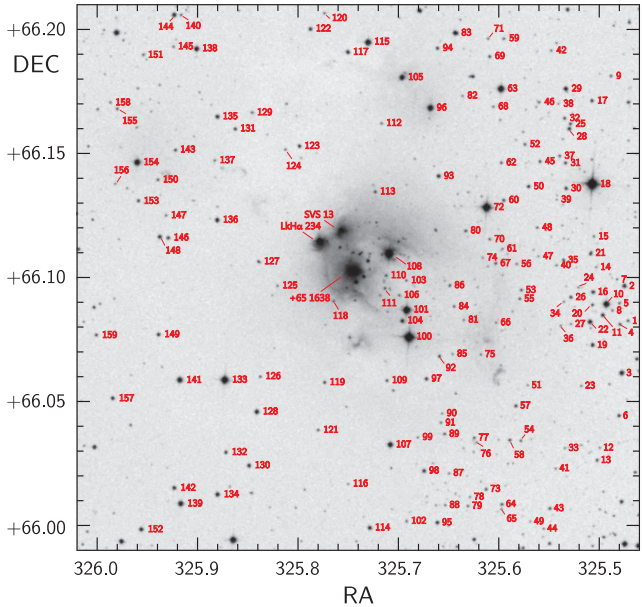}}
\vskip1mm
 \caption{Finding chart for the measured stars in the
NGC 7129 area. The star numbers are written on a DSS2 Red image
from SkyView.}
\label{3}
\end{figure}

Fig.\,1 shows the 13$\arcmin$\,$\times$\,13$\arcmin$ area with the NGC
7129 cluster in the centre (21$^{\rm h}$43$^{\rm m}$00$^{\rm s}$,
+66$\degr$06$\arcmin$).  In 2009 October this area was observed with the
1.8 m VATT telescope on Mt. Graham in Arizona with a CCD camera and
seven filters of the Vilnius photometric system.  The exposure lengths
were from 800 to 20 seconds in the two ultraviolet filters and from 100
to 4 seconds in the remaining filters.  The number of exposures used was
86.  The magnitudes determined from multiple exposures in the same
filter were averaged.  More details about the observations, their
processing and the classification of stars are given in Paper II.

In this area, 159 stars down to $V$ = 18.8 mag were measured.  The
distribution of the observed stars in apparent magnitudes is shown in
Fig.\,2.  Magnitudes and colours for the stars are listed in Table 1.
The columns list the following information:  star number, equatorial
coordinates J2000.0, magnitude $V$, colour indices $U$--$V$, $P$--$V$,
$X$--$V$, $Y$--$V$, $Z$--$V$ and $V$--$S$, photometric spectral types in
the MK system, and flags to the notes placed at the end of the table.
The rounded coordinates of stars were taken from the PPMXL catalog
\citep{Roeser2010}.  The finding chart for the measured stars in
the NGC 7129 area is given in Fig.\,3.

The r.m.s. errors of the magnitudes $V$ and colour indices $X$--$V$,
$Y$--$V$, $Z$--$V$ and $V$--$S$ to $V$ = 16 mag usually are lower than
0.02 mag, the errors of $U$--$V$ and $P$--$V$ are about 1.5--2.0 times
larger.  At $V$\,$\>$\,17.5 mag the accuracy of photometry in the
ultraviolet, especially for heavily reddened stars, is too low for a
reliable classification of stars.  For most of these stars the $U$-$V$
and $P$--$V$ colour indices are not given.  Colour indices with $\sigma$
= 0.05--0.10 mag are marked with colons. Most of these stars are either
fainter than 18 mag or are known as YSOs.

Single asterisks in the last column signify YSOs for which the notes are
given at the end of the table.  Most YSOs have been discovered  by
H$\alpha$ emission, by infrared excesses or by X-ray emission in the
papers described in the Introduction. Some YSOs were identified in the
present investigation using the $K_s$--[3.4] vs.  [3.4]--[4.6] diagram
which combines the 2MASS and WISE magnitudes.  The calibration of this
diagram in YSO classes was taken from \citet{Koenig2012}.  Variable
stars in the area were identified using the Catalogue of variable stars
in open cluster fields by \citet{Zejda2012}.  The stars found to be
binaries or having asymmetric images were not classified in luminosity
classes -- in the last column they are designated by double asterisks.


\begin{table*}
\def\hstrut{\vrule height10pt depth0pt width0pt}
\def\lstrut{\vrule height0pt depth6pt width0pt}
\caption{The extracted sample of the star data catalogue in the NGC 7129
area measured in the Vilnius seven-colour system.  The star numbers from
125 to 134 are taken to represent better all the columns.
The running numbers, coordinates, $V$ magnitudes, six colour indices,
photometric spectral types and notes are given.  The coordinates are
from the PPMXL catalogue \citep{Roeser2010} rounded to two decimals of
time second and to one decimal of arcsecond.  The full catalogue of 159
stars is available online.}
\label{table1}
\centering
\begin{tabular}{rrlclllllllc}
\hline\hline
 No. &  RA\,(J2000) & DEC\,(J2000) & $V$ & $U$--$V$ &
$P$--$V$ & $X$--$V$ & $Y$--$V$ & $Z$--$V$ & $V$--$S$ & Phot. &
Notes \hstrut \\
&  h~~~~m~~~~s~~~~ & ~~~$\circ$~~~~$\prime$~~~~$\prime\prime$ &
mag & mag  & mag & mag & mag & mag & mag  & sp. type &  \lstrut \\
\hline
\noalign{\vskip 0.5mm}
 125  &  21 43 16.83 &  +66 05 48.7 &  18.400 &         &         &  3.552: &  1.132 &  0.564: &  1.527  &  YSO        &   \z \\
 126  &  21 43 20.87 &  +66 03 37.0 &  17.487 &         &         &  2.702  &  1.078 &  0.605  &  1.249  &  m1 V       &    \\
 127  &  21 43 21.24 &  +66 06 23.8 &  17.456 &         &  3.944: &  2.984: &  1.203 &  0.531  &  1.426  &             &   \zz  \\
 128  &  21 43 21.70 &  +66 02 46.1 &  14.240 &  2.592  &  2.067  &  1.442  &  0.624 &  0.232  &  0.602  &  f8 V       &    \\
 129  &  21 43 22.90 &  +66 10 00.1 &  18.327 &         &         &  3.079: &  1.346 &  0.509: &  1.469  &  k0, YSO    &   \z \\
 130  &  21 43 23.54 &  +66 01 27.9 &  15.542 &  4.392  &  3.817  &  2.692  &  1.014 &  0.615  &  1.108  &  m0 V       &    \\
 131  &  21 43 26.95 &  +66 09 36.6 &  16.542 &  4.114  &  3.636  &  2.537  &  1.024 &  0.519  &  1.069  &  k3:, YSO    &   \z \\
 132  &  21 43 29.01 &  +66 01 47.0 &  16.094 &  3.802  &  3.311  &  2.250  &  0.839 &  0.444  &  0.894  &  k3 V       &    \\
 133  &  21 43 29.34 &  +66 03 31.9 &  11.124 &  2.339  &  1.782  &  1.184  &  0.516 &  0.181  &  0.497  &  f5 V, YSO:  &   \z \\
 134  &  21 43 31.01 &  +66 00 45.9 &  14.474 &  2.679  &  2.170  &  1.494  &  0.638 &  0.238  &  0.623  &  g0 V       &    \\
\hline
\end{tabular}
\end{table*}

\section{Interstellar reddening law and photometric classification}

From the Vilnius photometric data, spectral and luminosity classes in
the MK system were determined for 72 stars (Table 1 and Fig.\,2).  The
classification method was based on the interstellar reddening-free
$Q$-parameters and intrinsic colour indices, more information is given
in Papers I and II.  In Table 1 spectral classes are designated in the
lower-case letters to indicate that they are determined from photometric
data.  In calculating interstellar reddening-free $Q$-parameters and in
dereddening colour indices, the normal interstellar reddening law was
applied.  Its normality was verified with the $J$--$H$ vs.\,$H$--$K_s$
diagram (Fig.\,4) plotted for the NGC 7129 area which contains stars
with very different reddenings.


\begin{figure}
\resizebox{\hsize}{!}{\includegraphics{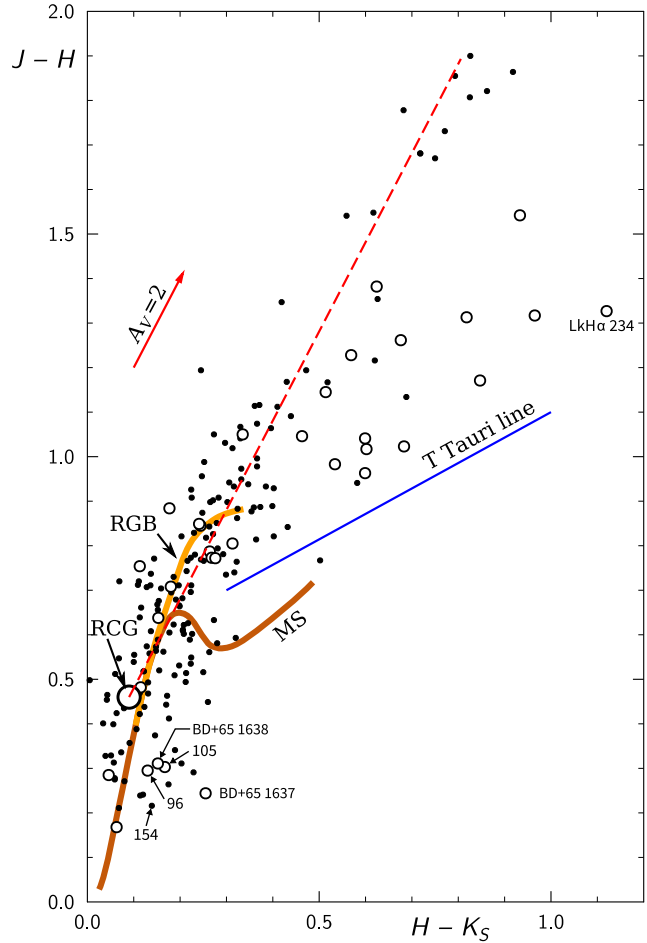}}
\vskip1mm
\caption{The $J$--$H$ vs.\,$H$--$K_s$ diagram for 2MASS stars with
errors of magnitudes $\leq$\,0.05 mag in the
15$\arcmin$\,$\times$\,15$\arcmin$ box centered on NGC 7129.  Above
$J$--$H$ = 1.5 the lower accuracy limit of magnitudes, $<$\,0.1, is
accepted.  The main sequence (MS, brown belt), red giant branch (RGB,
orange belt), the intrinsic locus of red clump giants (RCG) with its
reddening line, and the intrinsic T Tauri line are shown.  Open circles
designate the YSOs from Table 1 and from the literature.  The
six B3--A1 stars, used in Section 6 for the age determination, are
identified.}
\label{4}
\end{figure}

The area, from which the stars are plotted in Fig.\,4, is of the
15$\arcmin$\,$\times$\,15$\arcmin$ size, with the center on NGC 7129.
Lower than $J$--$H$ = 1.5, only the stars with magnitude errors $\leq
0.05$ mag are plotted.  The location of intrinsic sequences of
luminosity V and III stars and the intrinsic position of the Red Clump
Giants (RCGs) are shown according to \citet{StraizysLazauskaite2009}.
Trying to plot more heavily reddened giants, the magnitude errors above
the mentioned value of $J$--$H$ were increased up to 0.1 mag.
Most of the heavily reddened giants should be core helium-burning RCGs
since their space density outnumbers the density of normal
hydrogen-burning giants of spectral classes K--M by a factor of 10
\citep{Perryman1995, Perryman1997, Alves2000}.

The reddening line with the slope $E_{J-H} / E_{H-K_s}$ = 2.0, drawn
through the intrinsic position of RCGs at $J$--$H$ = 0.46, $H$--$K_s$ =
0.09
\citep{StraizysLazauskaite2009}, represents well the reddened giants up
to $J$--$H$ = 1.9.  This slope of the reddening line is typical of most
dark clouds and correspond to the normal interstellar reddening law
\citep{Straizys2008, StraizysLaugalys2008}.

The giants with $J$--$H$\,$>$\,1.5 are listed in Table 2.
Accepting that all these
stars are RCGs, we calculated for them the extinctions $A_{K_s}$ and
$A_V$, and distances with the equations
\begin{equation}
A_{K_s} = 2.0 E_{H-K_{s}} = 2.0 [(H-K_s) - 0.09],
\end{equation}
\begin{equation}
5\,\log d = K_s - M_{K_s} + 5 - A_{K_s},
\end{equation}
\begin{equation}
A_V = 8.3\,A_{K_s},
\end{equation}
with $M_{K_s}$ = --1.6 and $d$ in pc.


\begin{table*}
\def\hstrut{\vrule height10pt depth0pt width0pt}
\def\lstrut{\vrule height0pt depth6pt width0pt}
\caption{Probable red clump giants behind the TGU H645 P2
cloud identified in the $J$--$H$ vs. $H$--$K_s$ diagram.}
\label{table1}
\centering
\begin{tabular}{cccccrc}
\hline\hline
2MASS & $K$ & $J$--$H$ & $H$--$K_s$ & $A_{K_s}$ & $A_V$~ & $d$\,(kpc) \\
\hline
\noalign{\vskip 0.5mm}
J21424991+6602358 & 13.508 & 1.900 & 0.826 & 1.47 & 12.90 & 5.34 \\
J21431052+6601456 & 13.317 & 1.855 & 0.793 & 1.41 & 11.70 & 5.03 \\
J21425832+6607262 & 11.906 & 1.681 & 0.718 & 1.26 & 10.46 & 2.81 \\
J21430784+6607185 & 13.936 & 1.864 & 0.918 & 1.66 & 13.74 & 5.96 \\
J21431867+6604129 & 13.248 & 1.541 & 0.559 & 0.94 &  7.78 & 6.05 \\
J21432088+6606519 & 13.022 & 1.731 & 0.771 & 1.36 & 11.30 & 4.49 \\
J21425343+6611054 & 13.066 & 1.807 & 0.825 & 1.47 & 12.20 & 4.36 \\
J21425210+6607396 & 14.564 & 1.821 & 0.862 & 1.54 & 12.82 & 8.41 \\
J21425982+6607315 & 13.239 & 1.548 & 0.617 & 1.05 &  8.75 & 5.72 \\
J21431528+6607571 & 12.588 & 1.670 & 0.750 & 1.32 & 10.96 & 3.45 \\
J21431694+6612092 & 13.588 & 1.778 & 0.682 & 1.18 &  9.83 & 6.33 \\
\hline
\end{tabular}
\end{table*}

\section{Interstellar extinction}

Since the cluster NGC 7129 is embedded in the dust cloud, for the
determination of the cluster distance we have applied the method
which uses the rise of extinction of the field stars located at a
distance of the dust cloud.

Interstellar extinctions were calculated with the equation
\begin{equation}
A_V = 4.16~[(Y-V)_{\rm obs} - (Y-V)_0],
\end{equation}
where the intrinsic colour indices $(Y-V)_0$
for different spectral and luminosity classes were taken from
\citet{Straizys1992}. Distances to stars were calculated with the
equation
\begin{equation}
5 \log d = V - M_V +5 - A_V,
\end{equation}
the absolute magnitudes $M_V$ were taken from \citet{Straizys1992}, with
a correction of --0.1 mag, adjusting the $M_V$ scale to the
distance modulus of the Hyades, $V-M_V$ = 3.3 mag \citep{Perryman1998}.


\begin{figure}
\resizebox{\hsize}{!}{\includegraphics{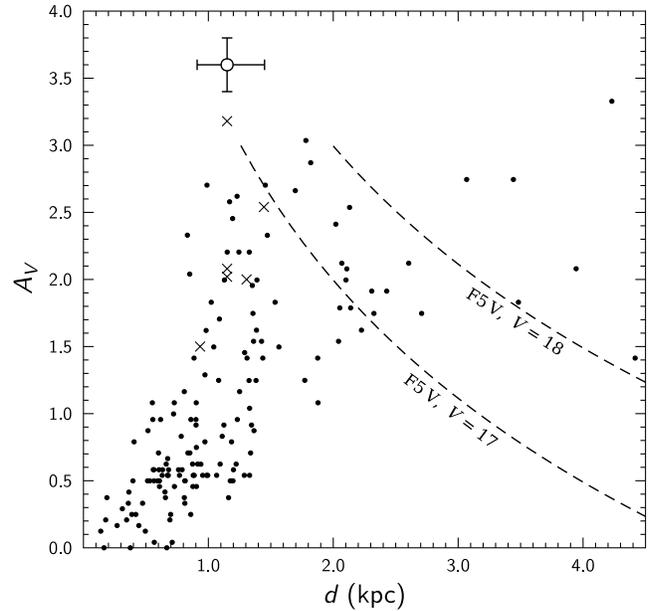}}
\vskip1mm
\caption{Extinction vs. distance diagram for 155 stars in
the 20$\arcmin$\,$\times$\,20$\arcmin$ area around NGC 7129 classified
in the Vilnius system. The crosses designate six cluster stars of the
highest luminosity.
The open circle with 3$\sigma$ error bars is plotted at the cluster
distance. The two curves show the limiting magnitude effect for
F5\,V stars with $V$ = 17 and 18 mag.}
\label{5}
\end{figure}

For the determination of the extinction run with distance in the
direction of NGC 7129, 72 stars from Table 1 with the most accurate
spectral types were used.  This sample was supplemented with 83 stars
classified with good accuracy in Paper I and located outside the
13$\arcmin$\,$\times$\,13$\arcmin$ area but inside the
20$\arcmin$\,$\times$\,20$\arcmin$ area centered on the cluster.  This
area covers the whole molecular cloud around NGC 7129 shown in the CO
map by \citet{Ridge2003}.  The known YSOs and other
stars with low classification accuracy were excluded.

Fig.\,5 shows the extinctions $A_V$ plotted as a function of the
distance $d$.  The crosses designate the six cluster members of the
highest luminosity.  These stars will be discussed in Section 6.  The
error bar in distance corresponds to $\Delta M_V$ = $\pm$\,0.5 mag, a
typical 3$\sigma$ error of absolute magnitudes estimated from
photometric spectral types.  The errors of $A_V$, originating from the
observational errors and intrinsic `cosmic dispersion' of the relation
between $Y$--$V$ and spectral classes, are of the order of $\pm$\,0.2
mag.

\section{Cloud distance}

The distribution of stars in Fig.\,5 shows that the extinction
increases steeply close to $d$\,$\approx$\,1 kpc where
the TGU H645 P2 cloud can be located.  The scatter of $A_V$ at greater
distances is quite large -- from about 1 mag to 3.4 mag.  Another, much
lower rise of extinction up to 1 mag might be present at a distance of
$\sim$\,500 pc.

For determining the cloud distance we must take into account that a
portion of stars are scattered toward lower distances because of
negative distance errors.  The main source of distance errors is in
their absolute magnitudes -- in the photometric classification we take
$\pm$\,0.5 as the 3$\sigma$ error of $M_V$.  In this case the stars with
the maximum negative distance errors will appear closer to the Sun by a
factor of 1.26.  Thus, if we find stars with large extinctions at a
distance $d_1$ pc, the true distance of the cloud should be at $d$ =
1.26\,$d_1$.  In Fig.\,5 the mean distance of five considerably
reddened stars ($A_V$\,$>$\,1.4 mag) at $d < 1.0$ kpc is 0.91 kpc.  If
these stars are moved from the dust cloud shortward because of absolute
magnitude errors, the front edge of the cloud is expected to be located
at $d$ = 1.15 kpc, which corresponds to the true distance modulus
$V$--$M_V$ = 10.30\,$\pm$\,0.17 mag.  This r.m.s. error of $V$--$M_V$
corresponds to the distance errors from --84 pc to +91 pc.  At a
distance of 1.15 kpc the angular diameter of the cloud, 0.3$\degr$, is
equal to 6 pc.

Also, for the shortward scattering of apparent distances the unresolved
binary stars can be responsible.  If both components of a binary star
are of the same luminosity, its real distance should be at 1.41\,$d_1$.
In this case the distance of the cloud should be at $d$ = 1.28 kpc.
Since we have no information that any of the five stars with apparent
distances $<$\,1 kpc is a binary, we will accept that their shifts
shortward are due to the errors of $M_V$ only.

As shown in Fig.\,2, most stars in the NGC 7129 area with
two-dimensional spectral types fall on 17--18 mag.  The majority of them
are main-sequence stars of spectral classes F and G. The two broken
curves in Fig.\,5 demonstrate the effect of limiting magnitude for F5\,V
stars with magnitudes $V$ at 17 and 18.  Above the upper curve only B-
and A-type stars, as well as G-K-M giants can be found.  These types of
stars in this area are rare.

Fig.\,5 shows that behind the dust cloud at 1.15 kpc the extinction does
not increase -- it remains approximately between 1.5 and 3.4 mag up to
4.5 kpc.  This is expected, since at the Galactic latitude 9$\degr$
our line of sight at 2 kpc reaches 320 pc above the Galactic plane,
where dust clouds are quite rare.  However, the $J$--$H$ vs.  $H$--$K_s$
diagram of 2MASS shows (Fig.\,4 and Table 2) that the background RCGs,
located at distances from 3 to 8 kpc, are affected by reddening, which
corresponds to $A_V$ up to $\sim$\,14 mag.  The most reddened 11 RCGs
with $J$--$H$\,$>$\,1.5 all are seen mostly in the eastern (left) half
of the cloud area with the largest dust density.  The presence among
them of
a few ordinary K--M giants will not change the conclusion that the total
extinction, created by the cloud in some directions, is as large as
10--14 mag.

\section{HR diagram}


\begin{figure}
\resizebox{\hsize}{!}{\includegraphics{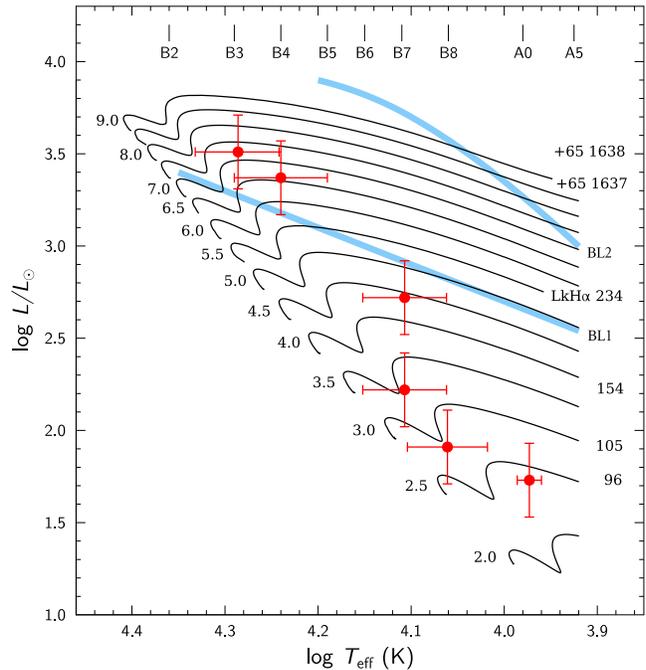}}
\vskip1mm
\caption{The effective temperature vs. luminosity diagram with the
pre-main-sequence evolution tracks for the masses 2--9
$M_{\odot}$ computed by the Pisa group. Six NGC 7129
members
of spectral classes B3--A1 located close to the ZAMS line with their
error bars are plotted in red.  Numbers of these stars are shown at the
right edge of the plot.  Spectral classes, corresponding to the
effective temperatures, are indicated at the top.  The two birthlines
for the 10$^{-5}$ and 10$^{-4}$\,$M_{\odot}$\,yr$^{-1}$ accretion rates
from \citet{Palla2005} are shown in blue.  They are designated as BL1
and BL2, respectively.}
\label{6}
\end{figure}

\begin{table*}
\begin{minipage}{140mm}
\def\hstrut{\vrule height10pt depth0pt width0pt}
\def\lstrut{\vrule height0pt depth6pt width0pt}
\caption{Photometric and evolution parameters of the most luminous stars
in the NGC 7129 area, located in the
luminosity-temperature diagram close to the main sequence. The results
for the first three stars are based on their MK types and $B,V$
photometry from \citet{Racine1968}, \citet{Herbst1999} and
\citet{Hernandez2004}, see the text. The results for the last three
stars are based on Vilnius photometry.}
\label{table1}
\begin{tabular}{llrrcccrccc}
\hline\hline
 No. &  Sp. type & $V$~~ & $M_V$ & $A_V$ & $\log L/L_{\odot}$ &
log\,$T_{\rm eff}$ & $BC$~ & Mass & Age & Notes \hstrut \\
&  &  & mag & mag &  &  & mag & $M_{\odot}$ & Myr &   \lstrut \\
\hline
\noalign{\vskip0.5mm}
BD+65 1637 & B4 (III-IV) & 10.18\rlap{v} & --2.14 & 2.02  &
3.37 & 4.240 & --1.57 & 6.6 & 0.27 & 1 \\
BD+65 1638 & B3 (IV)   & 10.18  & --2.20 & 2.08 &  3.51 &
4.286 & --1.85 & 7.2 & 0.20 &   \\
LkH$\alpha$\,234 & B7 (III) & 12.21\rlap{v} & --1.27 & 3.18 &
2.72 & 4.107 & --0.80 & 4.5 & 0.70 & 2  \\
2MASS J21424031+6610069  & a1 IV-V  & 12.35 & +0.80 & 1.50 &  1.73 &
3.973 & --0.16 & 2.4 & 3.10 &  3  \\
2MASS J21424707+6610512  & b8 V     & 13.34 &  0.00 & 2.54 & 1.91 &
4.061 & --0.55 & 3.0 & 2.38 &  4  \\
2MASS J21435035+6608477  & b7 V     & 12.28 & --0.30 & 2.00 & 2.22 &
4.107 & --0.80 & 3.5 & 1.59 &  5  \\
\hline
\end{tabular}
\vskip1mm
{\bf Notes.}\\
1. Herbig Ae/Be star; measured in the Vilnius system in Paper I (No.
663) and in Table 1 (No. 108).\\
2. Herbig Ae/Be star; measured in the Vilnius system in Paper I (No.
699). \\
3. No. 96 in Table 1, YSO of Class III, X-ray source
\citep{Stelzer2009}.\\
4. No. 105 in Table 1. \\
5. No. 154 in Table 1.
\end{minipage}
\end{table*}

Since the cluster NGC 7129 is embedded in the dust cloud TGU H645 P2, we
accept the same distance both for the cloud and the cluster.  We have
identified only six stars close to the ZAMS for which the membership to
the complex is evidenced by illumination of the surrounding dust and
forming the reflection nebulae.  Three of these stars, Nos. 96, 105 and
154, are 4--6$\arcmin$ away from the central concentration of visible
and infrared objects.  However, they all are located in the same dust
and molecular cloud, and we will consider these three stars as the
cluster members formed in the local condensations of gas and dust a few
million years ago.  The most active star formation continues up to now
in the core of the cluster which, according to \citet{Gutermuth2004,
Gutermuth2005}, has a diameter of 3$\arcmin$, while the whole area in
which YSOs are observed is about four times larger.  Our further
study of the cluster will be based only on these six stars which are
massive enough to be located in the temperature vs. luminosity diagram
close to the main sequence.  We recognize them as cluster members only
because they illuminate the surrounding dust cloud or are Herbig Ae/Be
stars.  Other cluster members are low-mass YSOs which have been easily
recognized by their emission lines, infrared excesses and X-ray emission
(see the Introduction).  However, within 6$\arcmin$ from the cluster
center we have not found more A or F stars whose distances would be
close to 1.15 kpc.  This means that during the last 1--3 Myr no more
massive stars have been formed.

The data for the six cluster stars are given in Table 3. Fig.\,6 shows
the plot of these stars in the the $\log L/L_{\odot}$ vs.  $\log T_{\rm
eff}$ diagram.  Their luminosities in solar units were calculated with
the equation
\begin{equation}
\log L/L_{\odot} = 0.4 (M_{\rm bol,\odot} - M_{\rm bol,\star}) =
0.4 (4.72 - V + A_V + 10.30 - BC),
\end{equation}
where $V$ is the apparent magnitude of the star, $M_{\rm bol,\star}$ is
its absolute bolometric magnitude, $M_{\rm bol,\odot}$ = 4.72 is the
bolometric absolute magnitude of the Sun, $BC$ are the bolometric
corrections, and 10.30 is the true distance modulus of the cluster.
The extinctions $A_V$ were determined with Eq. (4).

The three brightest stars in the cluster are BD+65 1637 (Herbig Ae/Be
star), BD+65 1638 (YSO of Class III) and LkH$\alpha$\,234 (also Herbig
Ae/Be star).  To calculate the $V_0$ = $V$ -- $A_V$ and ($Y$--$V$)$_0$
values of these stars, the following spectral classes were used:  B3 for
BD+65 1638 from \citet{Racine1968}, B4 for BD+65 1637 and B7 for
LkH$\alpha$\,234, both from \citet{Hernandez2004}.  The last two stars
are variables (V361 Cep and V373 Cep) with the $V$ amplitudes 0.71 and
1.48 mag, respectively.  Their positions in Fig.\,6 were calculated
using the average $V$ magnitudes and $B$--$V$ colours from
\citet{Herbst1999}.  For BD+65 1638 the extinction was calculated from
{\it BV} photometry given by \citet{Racine1968}.  If we place these
three stars at the accepted distance of the cloud (1.15 kpc), their
absolute magnitudes and corresponding luminosity classes are the
following:
--2.20 (B3\,IV) for BD+65 1638, --2.14 (B4\,III-IV) for BD+65 1637, and
--1.27 mag (B7\,III) for LkH$\alpha$\,234.  In Table 3 these luminosity
classes are given in parentheses.  In the $A_V$ vs.  $d$ diagram
(Fig.\,5) these three stars are plotted as crosses at the cluster
distance (1.15 kpc).

The three fainter stars, No.\,96 (2MASS J21424031+6610069, A1\,IV-V),
No.\,105 (2MASS J21424707+6610512, B8\,V) and No.\,154 (2MASS
J21435035+6608477, B7\,V), plotted in Fig.\,6, are the sources
illuminating the dust cloud in their vicinities and forming the three
small reflection nebulae around them.  The first two stars were observed
in the {\it JHKL} system by \citet{Strom1976} (SVS\,2 and SVS\,10).  The
reflection nebulae around them are described by \citet{Magakian1997}.
The star No.\,96 was identified by \citet{Stelzer2009} as YSO with X-ray
emission.  The stars Nos. 96 and 105 show excess emission in the WISE 12
and 22 $\mu$m bands.  However, both of them have approximately
photospheric spectral energy distributions up to $\lambda$ = 5--8
$\mu$m.  A possibility exists that the flux measurements of these stars
in the longest WISE bands are affected by a strong background of the
surrounding nebula (see the WISE images of NGC 7129 in SkyView and the
discussion in \citet{Koenig2012}).

The bolometric corrections and temperatures of stars were taken from
\citet{Straizys1992} according to their spectral classes. Our scale
of $T_{\rm eff}$ for B-stars is close to those given by
\citet{Flower1996}, \citet{Bessell1998} and \citet{Torres2010}.

We used PMS tracks in the mass range 2--9 $M_{\odot}$ adopting an
initial metallicity $Z$ = 0.0129 and helium abundance $Y$ = 0.274.
These values correspond to the initial [Fe/H] = 0.0, if the recent
photospheric ($Z$/$X$)$_{\odot}$ = 0.0181 from \citet{Asplund2009}, the
helium-to-metal enrichment ratio $\Delta Y/ \Delta Z$ = 2
\citep{Casagrande2007}, and the primordial helium abundance $Y_p$ =
0.2485 \citep{Cyburt2004} are adopted (see, e.g., equations 1 and 2 in
\citet{Tognelli2012} for details).  The models have been computed by the
Pisa group with the FRANEC stellar evolutionary code (see, e.g.,
\citet{Innocenti2008}), with input physics similar to those discussed in
detail in \citet{Tognelli2011} and used to compute the PMS tracks
currently available in the Pisa Stellar Models
Database\footnote{~http://astro.df.unipi.it/stellar-models/}.  The main
differences with respect to those PMS tracks are the adoption of
\citet{Asplund2009} solar heavy-element mixture instead of the
\citet{Asplund2005} one, and the consequent slightly higher solar
calibrated mixing length parameter value $\alpha$ = 1.74.  Another
important difference is the adoption of a mild (i.e.  $\beta_{ov}$ =
0.2) convective core overshooting in stellar models with $M \ge 1.2
M_{\odot}$.

The positions of the six stars with respect to the evolution tracks
allow to estimate their masses and ages, the results are given in Table
3. The ages of stars were read out from the evolution track tables at
the nearest positions.  Among the six stars, BD+65 1638 has the largest
mass (about 7.2\,$M_{\odot}$) and is the youngest (0.2 Myr).  The star
No. 96 has the smallest mass (2.4\,$M_{\odot}$) and it is the oldest one
(3.1 Myr).

It is worth to emphasize that the current generation of PMS models is
still quite uncertain for ages younger than about 1 Myr (see e.g.
\citet{Baraffe2002, Tognelli2011} and the discussion in the next
section). Thus, one should be cautious in assigning very young ages to
PMS stars.

The error crosses for each star are shown.  The errors of $\log
L/L_{\odot}$ in Fig.\,6 correspond to $\pm$\,0.5 mag absolute errors of
the accepted distance modulus of the cluster.  The errors log\,$T_{\rm
eff}$ correspond to an error of $\pm$\,1 decimal spectral subclass.
Naturally, for the two Ae/Be stars the real errors should be larger due
to lower accuracy of spectral classification, variability, presence of
circumstellar disks and envelopes, possibility of anomalous extinction
law in the circumstellar dust, etc.  The mass and age of a star would be
wrong if it were a binary.  However, we have not found in the literature
any indication on a possible duplicity of the six stars plotted in
Fig.\,6.  The vertical error bars of the distance modulus give various
mass errors for different stars -- from zero to one solar mass.

\section{The infrared group [BDS2003]31}

About a decade ago \citet{Bica2003b} have published a list of star
groups which in the atlas of the infrared 2MASS survey look like open
clusters.  One of these `infrared groups', [BDS2003]\,31, is located
close to NGC 7129, with the center coordinates RA = 21$^{\rm h}$42$^{\rm
m}$00$^{\rm s}$, DEC = +66$\degr$05$\arcmin$12$\arcsec$.  In Fig.\,1
this group of stars is surrounded by a 2$\arcmin$\,$\times$\,2$\arcmin$
square.

It is evident that this group is not an infrared object since
its 12 stars, seen in the $K_s$ filter, are all observable in optical
wavelengths.  Most of these stars were measured in the Vilnius system
(Table 1) and classified in two dimensions.  We plotted for the group
the $A_V$ vs.  $d$ and the $V_0$ vs.  ($Y$--$V$)$_0$ diagrams.  Both
these plots do not confirm that these stars form a real cluster:  their
$A_V$ are scattered within 0.3--1.8 mag, distances are within 0.4--5
kpc, and in the colour-magnitude diagram no sequence is seen.  One of
these stars, No.\,27 in Table 1, in the 2MASS + WISE diagram
$K_s$--[3.4] vs.  [3.4]--[4.6] looks like YSO.  Consequently, the object
[BDS2003]\,31 is just an accidental asterism, i.e., a group of unrelated
stars.

\section{Discussion}

In the literature, there are a few estimates of the age of the NGC 7129
stars based on different methods.  One of the methods is the estimation
of the fraction of stars with the circumstellar disks (YSOs of class II)
among the total cluster members.  According to \citet{Gutermuth2004}
this fraction is 54\%, while \citet{Stelzer2009} find 33\%.  Comparing
these fractions with the results for other young clusters, they estimate
the age of NGC 7129 as 2--3 Myr and 3 Myr, respectively.

\citet{Hernandez2004}, among other Herbig Ae/Be stars, have obtained new
estimates of spectral classes for BD+65 1637 and LkH$\alpha$\,234.
Combining these spectral classes with photometry from the literature,
they found the positions of these stars in the $\log L/L_{\odot}$
vs.  $\log T_{\rm eff}$ diagram and, comparing with the
PMS evolution tracks, estimated their masses and ages. The masses of
these stars are 7.0 and 4.8\,$M_{\odot}$, and the ages are
are 0.29 and 0.83 Myr, respectively, in close agreement with
our results, see Table 3.

\citet{Kun2009}, using their spectral classes and {\it BVRI} photometry
in NGC 7129, have plotted in the $\log L/L_{\odot}$ vs.  $\log T_{\rm
eff}$ diagram four NGC 7129 K--M stars of low masses.  Three of them are
found to be younger than 1 Myr. Increasing the distance from their 0.8
kpc to our 1.15 kpc leads to increase of the age but insignificantly.
However, the observed magnitudes, colour indices and bolometric
corrections of YSOs are usually affected by emission lines and infrared
excesses, therefore the calculation of their positions in the
theoretical HR plane can be inaccurate.  This was one of the
reasons why in the present paper we analysed only the massive stars
close to the main sequence:  four B-A stars with normal spectra and two
Herbig Ae/Be stars.

According to the concept of \citet{Stahler1983} (see also
\citet{Stahler2005, Stahler2013} and \citet{Palla2005}), all PMS tracks
in the HR diagram should begin from a `birthline' where the new-born
stars first appear as visible objects.  In the region of masses larger
than 1\,$M_{\odot}$ this line runs approximately along the 0.5--1.0 Myr
isochrones approaching the ZAMS.  The birthline intersects ZAMS at about
7\,$M_{\odot}$ for the accretion rate 10$^{-5}$\,$M_{\odot}$\,yr$^{-1}$
and at about 15\,$M_{\odot}$ for the accretion rate
10$^{-4}$\,$M_{\odot}$\,yr$^{-1}$.  Both these birthlines, BL1 and BL2,
are plotted in Fig.\,6 in blue.  The positions of BD+65 1637 and BD+65
1638 are close to the birthline BL1 but are slightly above it.  This
difference can be related to the errors in their luminosities and the
distance modulus of the cluster.  However, as recently shown by
\citet{Baraffe2012} the location of stars after the accretion phase is
still an open issue, being strongly affected by several not yet
well-constrained parameters (i.e., initial mass, accretion type,
accretion energy, see also \citet{Hosokawa2011}).  This might translate
in a quite large luminosity and effective temperature spread in the HR
diagram caused by the introduction of an early accretion phase, thus
making the comparison between data and models even harder.

\section{Results and conclusions}

1. Medium-band seven-colour photometry of 159 stars in the
13$\arcmin$\,$\times$\,13$\arcmin$ area in the direction of the cluster
and reflection nebula NGC 7129 in Cepheus is accomplished.

2. For 72 stars, using the interstellar reddening-free $Q$-parameters,
photometric spectral and luminosity classes in the MK system are
determined.

3. For these stars, supplemented with 83 stars from Paper I, the
interstellar extinction vs. distance diagram in the
20$\arcmin$\,$\times$\,20$\arcmin$ area centered on the cluster is
plotted.

4. The distance to the interstellar dust cloud TGU H645 P2, which
contains the embedded cluster, is found to be 1.15\,$\pm$\,0.08 kpc.
The extinction $A_V$ in the cluster area exhibits the values between 0.6
and 2.8 mag. The extinction in the densest parts of the cloud, estimated
from the $J$--$H$ vs. $H$--$K_s$ diagram, has the values up to
$\sim$\,13 mag.

5. For determining the age of NGC 7129, six cluster members of spectral
classes B3 to A1 were plotted in the $\log L/L_{\odot}$ vs.  $\log
T_{\rm eff}$ diagram, together with the Pisa evolution tracks.
Masses of the six stars are found between 7.2 and 2.4 $M_{\odot}$ and
the ages between 3.10 and 0.20 Myr.

6. We also conclude that the 'infrared group' [BDS2003]\,31, located
about 6$\arcmin$ from the core of NGC 7129, is not a physical cluster.

\section*{Acknowledgements} The use of the Simbad, WEBDA, ADS and
SkyView databases is acknowledged. We are grateful to
Francesco Palla for consultations.
The project is partly supported by the Research Council of Lithuania,
grant No.  MIP-061/2013.

\bibliographystyle{mn2e}
\bibliography{NGC7129}

\end{document}